\documentclass[preprint12pt]{elsarticle}
\usepackage[cp1251]{inputenc}
\biboptions{sort&compress}
\usepackage{graphicx}
\usepackage{bm}
\usepackage[T1]{fontenc}
\usepackage{amsmath}
\def\be{\begin{equation}}
\def\ee{\end{equation}}
\def\nn{\nonumber}
\def\d{{\rm d}}
\def\prl{\parallel}

\begin{document}
\begin{frontmatter}
\title{Angular momentum of radiation from ultrarelativistic charge}

\author[tspu,tsu]{Vladimir Epp\corref{mycorrespondingauthor}}
\cortext[mycorrespondingauthor]{Corresponding author}
\ead{epp@tspu.edu.ru}

\author[tspu]{Ulyana Guselnikova}
\ead{guselnikova.ulyana@yandex.ru}
 \affiliation[tspu]{organization={Tomsk State Pedagogical University},
             addressline={ul. Kievskaya 60},
             city={Tomsk},
             postcode={634061},
             country={Russia}}
\affiliation[tsu]{organization={Tomsk State  University},
             addressline={pr. Lenina 36},
             city={Tomsk},
             postcode={634050},
             country={Russia}}

\author[tpu]{Julia Janz}
\ead{yyg@tpu.ru}
 \affiliation[tpu]{organization={Tomsk Polytechnic University},
             addressline={pr. Lenina 30},
             city={Tomsk},
             postcode={634050},
             country={Russia}}
             
\begin{abstract}
The angular momentum of radiation from an arbitrarily moving relativistic charge is studied. The angular momentum is presented as the sum of the angular momentum relative to the point where the charge is located at a retarded moment of time and the angular momentum relative to an arbitrary stationary center. In particular, the instantaneous center of curvature of the trajectory is considered as such a center.
Explicit expressions for the angular distribution of these components of angular momentum fluxes are obtained and studied. It is shown that the angular momentum of the field relative to the position of the charge is determined only by the properties of the electromagnetic radiation field, and the angular momentum relative to an arbitrarily distant point is the vector product of the displacement of this point and the  force corresponding to radiation pressure.
It is shown that in the ultrarelativistic limit, the canonical angular momentum of the radiation coincides with the angular momentum following from the symmetrized energy-momentum tensor of the electromagnetic field.
\end{abstract}


\begin{keyword}
\texttt{vortex radiation\sep orbital angular momentum\sep charged particle \sep synchrotron radiation\sep relativistic}
\PACS[2010] 41.60.-m \sep 61.80.Fe \sep 42.50.Tx  
\end{keyword}

\end{frontmatter}

\section{Introduction}
Electromagnetic radiation can carry  angular momentum more than one Planck constant per photon. A characteristic feature of such radiation is a helical wave front.
A light beam with a  helical wave front is usually referred to as a twisted light.  In the optical range, twisted light is typically produced by lasers.
Methods for generating such laser radiation are described, for example, in  the  review \cite{Barnett2017} and books \cite{Torres_2011, Andrews2012book}. 

The vortex light beams  have opened a wide range of applications, such as spatial optical trapping of atoms or microscopic objects,   phase-contrast microscopy, and nano- or micro-scale physics \cite{Barnett2017,Andrews2012book}.
 Some authors proposed twisted radiation from rotating black holes or inhomogeneous interstellar media \cite{Tamburini_2011, Gray_2014} and even from single chiral molecule \cite{Hu_2024}. 

High energy photons carrying angular momentum can be   emitted by specially prepared vortex  beams of charged particles  \cite{Karlovets_2022, Zou_2024}.
Radiation in the X-ray range carrying the angular momentum can be obtained by converting an x-ray beam \cite{Kohmura2009} and by use of a helical undulator \cite{Bahrdt2013}. Various schemes of twisted photon beam production in undulators \cite{ Bordovitsyn2012,Matsuba_2018} and free electron lasers  \cite{Hemsing2011} have been proposed.
 Cherenkov radiation and transition radiation caused by vortex electrons was studied theoretically \cite{KarlovetsPhysRev, KarlovetsPhysRevLett}.
 Radiation of high energy charged particles  channeled in the solid and liquid crystals has been studied theoretically in recent papers \cite{Abdrashitov2018, EppJanz2018,Bogdanov_2021}. 
The  X-ray vortex radiation has found numerous applications both in classical and quantum optics condensed matter, high energy physics, optics, etc. (see  the review \cite{Hernandez-Garcia2017} and references therein).

In this work, we study the angular momentum flux in the radiation field of an arbitrarily moving ultrarelativistic charged particle. In particular, we investigate the case when the trajectory of the particle is sufficiently curved, so that the observer detects the radiation as a short flash emitted at the moment  when the particle's velocity is directed towards the observer. ``Sufficiently curved'' means that the force component $F_\perp$ transverse to the velocity satisfies the condition
 \cite{Landau_II} 
\be\label{con}
bF_\perp\gg mc^2,
\ee
where $b$ is  the size of the region in which the force changes noticeably, $m$ is the mass of the particle, $c$ is the speed of light. 
If this condition is met, the part of the trajectory from which the observer receives radiation can be approximated by a circular arc.

 The quantum probability of emission of a twisted photon by ultrarelativistic particles was studied in the recent paper \cite{Bogdanov2019PhysRev}. Here we examined this problem using the methods of classical electrodynamics.

The article is outlined as follows. In Section \ref{s2}, the angular momentum of the radiation field is presented as the sum of the ``intrinsic'' angular momentum relative to the point where the charged particle is located and the orbital angular momentum relative to an arbitrarily chosen point.  In Section \ref{s3}, the properties of the intrinsic angular momentum of the radiation are studied, and in Section \ref{s4}, the properties of the orbital angular momentum. In Section \ref{s5}, we integrate the angular momentum flux over time to find the total angular momentum passed per unit solid angle during a single passage of the point at which the particle's velocity is directed toward the observer.
Spectrum properties  of the angular momentum flux are studied in  Section \ref{s6}.

\section{``Intrinsic'' and ``orbital'' angular momentum}
\label{s2} 
 The angular momentum of the electromagnetic field, by definition, depends on the choice of coordinate system. The most studied is the angular momentum of radiation from particles moving in a circle or spiral \cite{Bahrdt2013, Bordovitsyn2012, Matsuba_2018, Bogdanov2018, EppGuselnikova2019}. The angular momentum of radiation is calculated relative to the center of the circle. In this case, it is the orbital momentum that makes a significant contribution to the angular momentum of the radiation pulse.  This is especially obvious for the radiation of an ultrarelativistic charge, when radiation is concentrated in a narrow cone in the direction of the particle's velocity. In particular, for synchrotron radiation of an electron in a cyclic accelerator, the angular momentum transferred to the target is, roughly speaking, the product of the radiation pressure multiplied by the radius of the accelerator. This is essentially the orbital angular momentum. However, of interest is the part of angular momentum that does not depend on how far we place the origin of coordinates from the radiation source, relative to which we calculate the angular momentum. We will call this part of the angular momentum the ``intrinsic'' angular momentum. The part proportional to the distance between the charged particle and coordinate origin we will denote as ``orbital'' angular momentum. In this section we will explicitly separate the intrinsic angular momentum of the radiation field from the orbital angular momentum.
 
 The angular momentum of the radiation field $\d\bm L$ incident on the area $\d s$ during the time $\d t$ is calculated through the Poynting vector  $\bm P$ \cite{Jackson_Cl_El, Panofsky1962}
 \be\label{Lds-nn}
\frac{\d\bm L}{\d s\,\d t}=\frac{1}{c} (\bm r\times \bm P),\quad \bm P=\frac{c}{4\pi}(\bm E\times \bm H).
\ee
Here $\bm r$ is a vector pointing from the origin of the coordinate system, relative to which we want to find the angular momentum, to the point where the observer is located, $\bm E$ and $\bm H$ are the electric and magnetic fields of the point charge, respectively
\cite{Landau_II}
\begin{align}\label{EH}
\bm E&=\bm E_1+\bm E_2,\quad \bm H=(\bm R\times \bm E)/R,\\
\label{ee}
\bm E_1&=\frac{eR^2\bm \kappa}{c(R-\bm \beta\bm R)^3},\quad
\bm E_2=\frac{e(1-\beta^2)(\bm R-R\bm\beta)}{(R-\bm \beta\bm R)^3},\\
\bm \kappa&=[\bm R\times[(\bm R-R\bm\beta)\times\bm{\dot{\beta}}]]/R^2,\nn
\end{align}
$\bm R$ is the vector connecting the position of the particle with the observer at the retarded  time $t'=t-r/c$, $c$ is the speed of light, $e$ is the charge of the particle, $\bm \beta=\bm v/c,\, \bm v=\bm v(t')$ is the particle velocity, the dot means the derivative with respect to time $t'$. The field $\bm E_1$ decreases with distance as $1/R$, and the field $\bm E_2$ as $1/R^2$.

 If we express the vector $\bm H$ in terms of $\bm E$, then the Poynting vector can be written as
 \[
 \bm P=\frac{c}{4\pi R}\left[\bm R E^2-\bm E(\bm R\bm E_2)\right].
 \] 
 In  the second term, we can leave only vector $\bm E_1$ in the vector $\bm E$, since $\bm E_2$ decreases as $1/R^2$
 \[
 \bm P=\frac{c}{4\pi R}\left[\bm R E^2-\bm E_1(\bm R\bm E_2)\right].
 \] 
 Thus, the Poynting vector in the wave zone is divided into two terms. The first, proportional to $\bm R$ is parallel to the direction of radiation. It, in particular, determines the intensity of radiation. The second term is orthogonal to the direction of radiation. 
 
Let us represent the radius vector of the observer $\bm r$ as the sum $\bm r=\bm R+\bm r'$, where $\bm r'$ is the radius vector of the charged particle at  retarded  time. Then the angular momentum flux (\ref{Lds-nn}) takes on the form
\be\label{Lds12}
\frac{\d\bm L}{\d s\,\d t}=\frac{\d\bm L_0}{\d s\,\d t}+\frac{\d\bm L'}{\d s\,\d t},
\ee
where 
\be\label{Lds1-2}
\frac{\d\bm L_0}{\d s\,\d t}=\frac{1}{c} (\bm R\times \bm P),\quad \frac{\d\bm L'}{\d s\,\d t}=\frac{1}{c} (\bm r'\times \bm P).
\ee
The angular momentum $\bm L_0$ is determined relative to the point of the charge location. It does not depend on the choice of coordinate system and is determined only by the speed and acceleration of the charged particle. We denote this part of the angular moment as intrinsic angular moment. The angular momentum $\bm L'$ is  proportional to the distance of the charge from the origin and in this sense it can be designated as the orbital angular momentum.

 Leaving only the largest terms in the expansion in powers of $1/R$, we obtain
 \be\label{Lvec}
\frac{\d\bm L_0}{\d s\d t}=\frac{1}{4\pi R}(\bm E_2\cdot\bm R)(\bm E_1\times\bm R),\quad 
\frac{\d\bm L'}{\d s\d t}=\frac{1}{4\pi R}E_1^2(\bm r'\times\bm R).
\ee
In the limit  $R\to\infty$ the first expression can be written as
 \be\label{Lvecc}
\frac{\d\bm L_0}{\d \Omega\d t}=\frac{e^2 (1-\beta^2)(\bm\kappa\times \bm n)}{4\pi c (1-\bm\beta\bm n)^5}.
\ee
Here $\d\Omega$ is the solid angle based on the area $\d s$, $\bm n=\bm R/R$ is the unit vector in the direction of radiation. Note that this expression includes both the field $\bm E_1$, which decreases with distance as $1/R$, and the field $\bm E_2$, which decreases according to the law $1/R^2$. 

Let us find the flux density of the vector $\bm L'$. In the limit $R\to\infty$, only the field $\bm E_1$ remains in the flux density
 \be\label{LE1}
\frac{\d\bm L'}{\d\Omega\d t}=\frac{1}{4\pi}E_1^2(\bm r'\times\bm n)=\frac{e^2\kappa^2(\bm r'\times\bm n)}{4\pi c^2 (1-\bm\beta\bm n)^6}.
\ee
Summing up equations  (\ref{Lvecc}) and (\ref{LE1}), we get the total flux dencity of angular momentum in the radiation field
\be\label{vece}
\frac{\d\bm L}{\d\Omega\d t}=\frac{e^2}{4\pi c (1-\bm\beta\bm n)^5}\left[ (1-\beta^2)(\bm\kappa\times \bm n)+\frac{\kappa^2(\bm r'\times\bm n)}{c(1-\bm\beta\bm n)}\right].
\ee
This formula was obtained in \cite{Epp_2022PR} using the symmetrized energy-momentum tensor of the electromagnetic field. The first term in (\ref{vece}) represents the angular momentum flux relative to the point at which the charge is located, the second term is the orbital angular momentum flux.

If we interpret the angular momentum mechanistically as  ``the product of linear momentum and lever arm'', then the first term has no mechanical analogy, since the ``lever arm'' in the case under consideration is zero. And this term does not represent the spin of photons since it is orthogonal to the direction of radiation. 
The second term in (\ref{vece}) has a direct mechanical analogy, since $\bm n E_1^2/4\pi$ (see (\ref{Lvec})) is the radiation pressure force per unit area, and the ``lever arm'' is represented by the vector $\bm r'$.

\section{Flux of the intrinsic angular momentum}\label{s3}
\begin{figure}[ht]\center
\includegraphics[width=11cm]{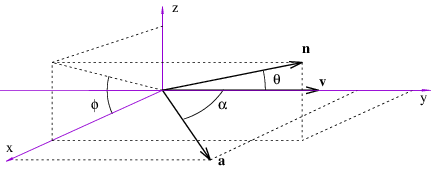}
\caption{Coordinate system.}
\label{sc}
\end{figure}

We choose a coordinate system so that the origin coincides with the position of the particle, the $y$ axis is directed along the particle velocity, and the particle acceleration lies in the $xy$ plane. The angle between the velocity vector $\bm v$ and acceleration vector $\bm a$ we denote by $\alpha$. We choose the direction of the $z$ axis so that  $(\bm a\times \bm v)_z>0$. We specify the direction of the unit vector $\bm n$ by the angles $\theta$ and $\phi$ as shown in Fig. \ref{sc}.
Since the vector $\bm L_0$ is orthogonal to the direction of radiation, it is convenient to express it via projections onto the  basis vectors of the spherical coordinate system $\hat{\bm\theta}$ and $\hat{\bm\phi}$. Their Cartesian coordinates read 
\begin{align}\label{bas}
\bm{\hat\theta}=&(\cos\theta\cos\phi,\,-\sin\theta,\, \cos\theta\sin\phi)\nn,\\
\bm{\hat\phi}=&(-\sin\phi,\, 0, \cos\phi).
\end{align}
Expanding the scalar and vector products in  (\ref{Lvecc}), we obtain
\begin{align}\label{Lsfer1}
\frac{\d L_{0\theta}}{\d\Omega\d t}=&-\frac{e^2\dot\beta}{4\pi c}\frac{(1-\beta^2)}{(1-\beta\cos\theta)^4}\sin\alpha\sin\phi,\\
\label{Lsfer2}
\frac{\d L_{0\phi}}{\d\Omega\d t}=&\frac{e^2\dot\beta}{4\pi c}\frac{(1-\beta^2)}{(1-\beta\cos\theta)^5}\left[ (\beta-\cos\theta)\cos\phi\sin\alpha+\sin\theta\cos\alpha\right].
\end{align}
If the acceleration is parallel to the velocity, the flux of angular momentum  is axially symmetric and has only the azimuthal component $L_{0\phi}$. 
 The sign of the azimuthal component is determined by the left-hand rule regarding the direction of acceleration. If the acceleration is in the opposite direction to the speed ($\alpha=\pi$), the angular momentum will have direction  opposite to the direction at $\alpha=0$.

Let us find asymptotic expressions for the flux of angular momentum in the radiation field of an ultrarelativistic particle. In this case $\gamma=(1-\beta^2)^{-1/2}\gg 1$ and $\theta\ll 1$. Since the main part of the radiation is concentrated in a narrow cone in the direction of velocity, it is convenient to study the projections of the vector $\bm L_0$ onto the plane $xz$ orthogonal to the velocity.
Let us introduce the notations
\[
\psi=\theta\gamma,\quad \psi_x=\psi\cos\phi,\quad \psi_z=\psi\sin\phi.
\]

The Cartesian coordinates of the vector $\bm \kappa$ included in the expression (\ref{Lvecc}) are in this approximation equal (in what follows we will put the equal sign instead of the $\approx$ sign)
\be\label{vecs1}
\bm\kappa=\frac{1}{2\gamma^2}\dot\beta\left[(\psi_x^2-\psi_z^2-1)\sin\alpha+2\gamma\psi_x\cos\alpha,\,-2\psi^2\cos\alpha,\,2\psi_z(\psi_x\sin\alpha+\gamma\cos\alpha)\right].
\ee
and the parenthesis in the denominator is equal to
\be\label{bn}
1=\bm n\bm\beta=\frac{1}{2\gamma^2}(1+\psi_x^2+\psi_z^2).
\ee
In these notations, the angular momentum flux $\bm L_0$ has the form
 \begin{align}\label{Lsfer1b}
\frac{\d L_{0x}}{\d\Omega\d t}=&-\frac{8 e^2\dot\beta\gamma^6\psi_z}{\pi c(1+\psi^2)^5}(\psi_x\sin\alpha+\gamma\cos\alpha),\\
\label{Lsfer2b}
\frac{\d L_{0y}}{\d\Omega\d t}=&\frac{4 e^2\dot\beta\gamma^5\psi_z\sin\alpha}{\pi c(1+\psi^2)^4},\\
\label{Lsfer3b}
\frac{\d L_{0z}}{\d\Omega\d t}=&\frac{4 e^2\dot\beta\gamma^6}{\pi c(1+\psi^2)^5}\left[ (\psi_x^2-\psi_z^2-1)\sin\alpha+2\gamma\psi_x\cos\alpha\right].
\end{align}

The terms with $\cos\alpha$ contain a large factor $\gamma$, but we do not neglect the other terms because $\cos\alpha$ can be small. The point is that in the ultrarelativistic case the angle between the velocity and acceleration of the particle is close to $\pi/2$, if the angle between the force acting on the particle and its velocity is of the order of unity. Indeed, expanding the force $\bm F$
\[
\bm F= \frac{\d}{\d t}\frac{m\bm v}{\sqrt{1-\beta^2}}=mc [\bm{\dot\beta}\gamma+\bm\beta(\bm\beta\bm{\dot\beta})\gamma^3]
\] 
into components longitudinal and transverse relative to the velocity   $F_\prl=\dot\beta_\prl\gamma^3$
and $F_\perp=\dot\beta_\perp\gamma$, we obtain
\[
\tan\alpha=\frac{\dot\beta_\perp}{\dot\beta_\prl}=\frac{F_\perp}{F_\prl}\gamma^{2}.
\]
Hence, $\tan\alpha$ is of order $\gamma^2$.

\begin{figure}[ht]\center
\includegraphics[width=5.7cm, height=5.4cm]{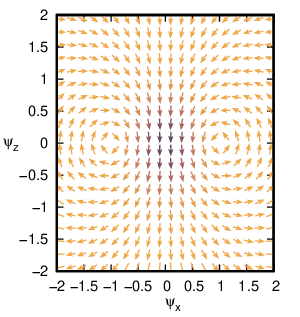}\quad
\includegraphics[width=5.7cm, height=5.4cm]{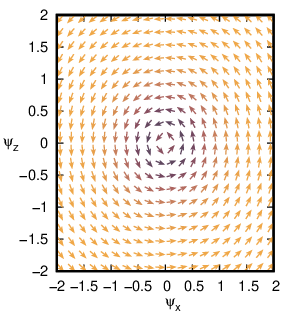}
\caption{Vector $\bm L_0$ in the $xz$ plane with $\alpha=\pi/2$ (left panel) and $\alpha=0$ (right panel).}
\label{vvec}
\end{figure}

Fig. \ref{vvec} shows the angular distribution of the angular momentum  flux in the $xz$ plane for two values of the angle $\alpha$. 
If the acceleration is orthogonal to the velocity, the main part of  radiation carries angular momentum orthogonal to the trajectory plane. In this case, there are two directions $\psi_x=\pm 1$, $\psi_z=0$, in which the angular momentum of  radiation is zero.
If the acceleration is parallel to the velocity, the radiation has only the azimuthal angular momentum component $L_{0\phi}$.  As noted above, the sign of the azimuthal component is determined by the left-hand rule regarding the direction of acceleration. 
If the acceleration is opposite to the velocity ($\alpha=\pi$), the angular momentum will be in the opposite direction to the direction shown in the right panel of Fig.  \ref{vvec}.

\begin{figure}[ht]\center
\includegraphics[width=6.4cm]{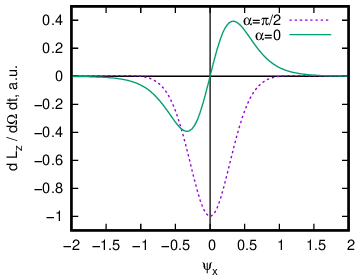}
\caption{The plot of $\d L_{0z}/\d\Omega\d t$ in the $\psi_z=0$ plane as a function of $\psi_x$. The graph for $\alpha=\pi/2$ is reduced by $\gamma$ times.}
\label{vvec1}
\end{figure}

The value of the $z$-component of angular momentum in the $z=0$ plane is shown in Fig.  \ref{vvec1}. 
The presented curves have different scales - the angular momentum flux at $\alpha=0$ is approximately $\gamma$ times greater than at 
$\alpha=\pi/2$ in accordance with  (\ref{Lsfer3b}).

\begin{figure}[ht]\center
\includegraphics[width=6.8cm]{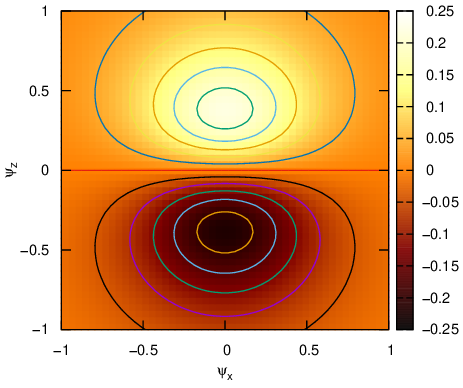}
\caption{Flux of the projection of the vector $\bm L_0$ onto  direction of velocity at $\alpha=\pi/2$.}
\label{L-v}
\end{figure}

Equation (\ref{Lsfer2b}) shows that the flux of the projection of the angular momentum onto the velocity direction is $\gamma$ times smaller in order of magnitude than the projection onto the plane orthogonal to the velocity. However, this component may be of interest because it is responsible for the torque that rotates the target on which the radiation falls about the direction of  radiation. The map of the flux of this longitudinal component  of angular momentum is shown in Fig. \ref{L-v}. This component has different signs above and below the orbital plane.

\section{Flux of orbital angular momentum of radiation}\label{s4}
The orbital angular momentum $\bm L'$ in the radiation field  depends on the point relative to which the angular momentum is calculated. Let us find the flux of angular momentum $\bm L'$ relative to the instantaneous center of curvature of the trajectory. The center of curvature lies on the normal to the trajectory in the trajectory plane. The radius of curvature $\rho$ is equal to the reciprocal of the curvature
\[
\rho=\frac{v^3}{|(\bm v\times\bm{\dot v})|}=\frac{v^2}{\dot v\sin\alpha}.
\]
The flux of orbital angular momentum relative to an arbitrary point is expressed by the formula (\ref{LE1}).
The vector $\bm r'$   is equal to
\[
\bm r'=-(\rho,\, 0,\, 0).
\]
If the velocity is parallel to the acceleration, then $\alpha=0$. The radius of the trajectory in this case is infinite and the flux of orbital angular momentum relative to the instantaneous center of curvature is also infinite.

It follows from  (\ref{LE1}) that for the arbitrary velocity of the charge,  the flux of orbital angular momentum in the case of $\alpha=\pi/2$ is
\be\label{Lsl1}
\frac{\d\bm L'}{\d\Omega\d t}=-\frac{e^2\beta^2\dot\beta}{4\pi c (1-\beta\cos\theta)^4}\left[1-\frac{(1-\beta^2)\sin^2\theta\cos^2\phi}{(1-\beta\cos\theta)^2}\right](\hat\theta\,\sin\phi+\hat\phi\cos\theta\cos\phi).
\ee
We will write the asymptotic expression for  $\gamma\gg 1$ in the Cartesian coordinate system similarly to the equations (\ref{Lsfer1b}) -- (\ref{Lsfer3b})
\begin{align}\label{Lsfer1c}
\frac{\d L'_{x}}{\d\Omega\d t}=&0,\\
\label{Lsfer2c}
\frac{\d L'_{y}}{\d\Omega\d t}=&-\frac{4 e^2\dot\beta\gamma^7\psi_z\left[(1+\psi^2)^2-4\psi_x^2 \right]}{\pi c(1+\psi^2)^6},\\
\label{Lsfer3c}
\frac{\d L'_{z}}{\d\Omega\d t}=&\frac{4 e^2\dot\beta\gamma^8\left[(1+\psi^2)^2-4\psi_x^2 \right]}{\pi c(1+\psi^2)^6}.
\end{align}
Comparison with  (\ref{Lsfer1b})-(\ref{Lsfer3b}) shows that the flux of orbital angular momentum is, by order of magnitude, $\gamma^2$ times greater than the flux of the intrinsic  angular momentum. The main contribution to the orbital angular momentum flux comes from the $\bm L'_z$ component orthogonal to the trajectory plane.

\section{Total angular momentum of radiation}\label{s5}
The dependence on the angle $\psi_x$  in the  above equations is not of interest, since when the radiation cone passes through the observer’s position, it almost uniformly covers a strip  lying in the plane of the particle’s orbit with vertical angular width of the order of $\gamma^{-1}$. Now we are interested in the dependence of the angular momentum flux on the angle $\psi_z$ in this strip.
To find this, we will proceed as follows \cite{Bayer1973, Hofmann}. We  assume that the condition (\ref{con}) is satisfied. We approximate the trajectory of the charge in the vicinity of the coordinates  origin by a circular arc. The dependence of the vectors $\bm r',\,\bm\beta$ and $\bm{\dot\beta}$ on the retarded time $t'$ is
 \begin{align}
\bm r'(t')=&\rho\left[-\cos\omega t',\, \sin\omega t', \, 0\right]\approx \rho\left[\frac{1}{2} (\omega t')^2-1,\, \omega t'-\frac 16  (\omega t')^3,\, 0\right],\\
\bm \beta\approx & \beta\left(\omega t',\,1-\frac 12 (\omega t')^2,\, 0\right),\\
\bm{\dot\beta}\approx &\dot\beta (1,\,\omega t',\,0),
\end{align}
where $\omega=\dot v/v$ and $\rho=v/\omega$ are the angular velocity and radius of the arc of trajectory. The condition (\ref{con}) implies that radiation in a given direction is generated during the time $\omega t'\sim \gamma^{-1}$. Let us find the angular distribution of the angular momentum flux in the $yz$ plane. Namely, we will assume that the vector $\bm n$ has coordinates
\[
\bm n= (0,\,\cos\theta,\,\sin\theta)\approx \left(0,\,1-\frac 12 \theta^2,\,\theta\right).
\]
For example, dependence of $1-\bm\beta\bm n $ on time $t'$ looks like
\be\label{vecs11}
1-\bm\beta\bm n = \frac{1}{2\gamma^2}\left(1+\psi_z^2+(\gamma\omega t')^2\right),
\ee
This expression differs from  (\ref{bn}) by replacing $\psi_x$ by $\gamma\omega t'$. This is a consequence of the fact that 
rotation of the coordinate system by an angle $\gamma^{-1}\psi_x$ around the $z$ axis is equivalent to a rotation of the radius vector $\bm r'$ by an angle $\omega t'$ due to particle motion.
Obviously, such a replacement must be made in all equations for the instantaneous angular distribution in order to obtain the dependence of the angular momentum flux as a function of time. Having made such a replacement, we integrate  (\ref{Lsfer1b})--(\ref{Lsfer3b}) and (\ref{Lsfer1c})--(\ref{Lsfer3c}) over time $t$. Herewith, we take into account that
\[
\d t=(1-\bm\beta\bm n)\d t' .
\]
Integrating (\ref{Lsfer1b})--(\ref{Lsfer3b}) we obtain
\begin{align}\label{Lsfer1d}
\frac{\d L_{0x}}{\d\Omega}=&0,\\
\label{Lsfer2d}
\frac{\d L_{0y}}{\d\Omega}=&\frac{5 e^2\gamma^2\psi_z}{8 c(1+\psi_z^2)^{7/2}},\\
\label{Lsfer3d}
\frac{\d L_{0z}}{\d\Omega}=&-\frac{e^2\gamma^3}{2 c(1+\psi_z^2)^{5/2}}.
\end{align}
These equations determine the angular distribution of angular momentum flux in the $yz$ plane. It is more convenient to take here as the infinitesimal  solid angle
\[
\d\Omega=\d\theta\d\varphi=\gamma^{-1}\d \psi_z\d\varphi,
\]
where $\varphi$ is the azimuthal angle in plane $xy$. 
 It follows from  (\ref{Lsfer2d}) that the projection of the angular momentum of the field onto  direction of the particle velocity is positive above the orbital plane ($\psi_z >0$) and negative below the orbital plane ($\psi_z <0$). The angular momentum component $L_{0z}$, which has passed into the solid angle $\d\Omega$ over the entire time, is $\gamma$ times greater than the $L_{0y}$ component.

Integrating the orbital angular momentum flux (\ref{Lsfer1c})--(\ref{Lsfer3c}) over time gives the total angular momentum emitted into the solid angle $\d\Omega$ for all time
\begin{align}\label{Lsfer1e}
\frac{\d L'_{x}}{\d\Omega}=&0,\\
\label{Lsfer2e}
\frac{\d L'_{y}}{\d\Omega}=&\frac{e^2\gamma^4\psi_z(7+12\psi_z^2)}{16 c(1+\psi_z^2)^{7/2}},\\
\label{Lsfer3e}
\frac{\d L'_{z}}{\d\Omega}=&-\frac{e^2\gamma^5(7+12\psi_z^2)}{16 c(1+\psi_z^2)^{7/2}}.
\end{align}

Finally, we integrate the angular distributions  (\ref{Lsfer1d})--(\ref{Lsfer3d}) and (\ref{Lsfer1e})--(\ref{Lsfer3e})  over the angle $\psi_z $.
As a result, we obtain the total angular momentum emitted into the azimuthal angle $\d\varphi$ lying in the plane  of the trajectory
\be\label{i21}
\frac{\d L_{0z}}{\d\varphi}=-\frac{e^2\gamma^2}{3c},\quad \frac{\d L'_{z}}{\d\varphi}=-\frac{2e^2\gamma^4}{3c}.
\ee
The remaining components of the vectors $\bm L_0$ and $\bm L'$ are equal to zero. The angular momentum $L_{0z}$ relative to the point of the charge location  is $\gamma^2$ times less than the angular momentum $L'_z$ relative to the instantaneous center of curvature.

Note that the result of integration over time does not depend on the acceleration of the charge and, therefore, on the curvature of the trajectory. Although, the instantaneous flux of angular momentum (\ref{Lsfer1b})--(\ref{Lsfer3b}) and (\ref{Lsfer1c})--(\ref{Lsfer3c}) is proportional to acceleration. This is due to the fact that the time during which an observer detects a flash of radiation is inversely proportional to the curvature of the trajectory.

We found the angular distribution of $\bm L'$ relative to the instantaneous center of curvature. If we calculate $\bm L'$ relative to an arbitrary point lying on the principal normal to the trajectory at a distance of $r'$ from the charge, then in accordance with (\ref{LE1}), $\bm L'$ will be $r'/\rho$ times greater ($\rho$ being the curvature radius).
\section{Spectral decomposition of angular momentum flux}\label{s6}
When calculating the spectral distribution of angular momentum, we  assume that the condition (\ref{con}) is satisfied.  We approximate the section of the trajectory from which radiation is emitted in a given direction by a circular arc and use the results obtained elsewere  for synchrotron radiation.

The spectral-angular distribution of the flux of angular momentum for synchrotron radiation was obtained by different methods and by several authors. For a particle moving in a circle with a frequency $\omega$, semiclassical \cite{Bogdanov2018} and quantum calculations \cite{Epp_2023} give in the classical limit 
\be\label{sch11}
\frac{\d L_{z}}{\d\Omega\d t}=\frac{e^2\omega}{2\pi c}\sum_{n=1}^\infty n \left[\beta^2 {J'_n}^2(\xi)+\cot^2\vartheta\, J^2_n(\xi)\right].
\ee
Here $J_n(\xi)$ and $J'_n(\xi)$ are the Bessel functions and their derivatives, respectively, $\xi=n\beta\sin\vartheta$, $\vartheta$ is the angle between the direction of radiation and the $z$ axis. The $z$ axis is orthogonal to the particle orbit, and its direction is chosen so that $(\bm\beta\times\bm{\dot\beta})_z>0$.

At the same time, representing  (\ref{Lds-nn}) as a Fourier series leads to a slightly different expression \cite{EppGuselnikova2019}
\be\label{sch12}
\frac{\d L_{z}}{\d\Omega\d t}=\frac{e^2\omega\beta\sin \vartheta}{2\pi c}\sum_{n=1}^\infty n\left\{ \xi\left[{J'_n}^2(\xi)+\cot^2\vartheta\, J^2_n(\xi)\right]
+J_n(\xi) J'_n(\xi)\right\}.
\ee
If the canonical angular momentum is used as the angular momentum of radiation, we again obtain the expression (\ref{sch11})  \cite{Epp_2023}.
 The canonical angular momentum, is defined as \cite{Jackson_Cl_El, Bliokh_2013, Afanasev_2022,Yang_2022}
\be\label{Lcan}
\frac{\d {\bm L}}{\d t}=cr^2\int\bm M\d\Omega,
\ee
where the vector $\bm M$ represents the sum of the ``orbital'' $\bm{\mathcal L}$ and ``spin'' $\bm{\mathcal S}$ angular momentum
\be\label{Mvec}
\bm M=\bm{\mathcal L}+\bm{\mathcal S},
\ee
 \be\label{MM}
 \bm {\mathcal L}=\frac{1}{4\pi c}E_i(\bm r\times\bm\nabla)A_i,\quad \bm {\mathcal S}= \frac{1}{4\pi c}(\bm E\times\bm A), 
 \ee 
and  vector $\bm A$ is the vector potential in the Coulomb gauge $\bm \nabla\bm A=0$.

The main difference between  (\ref{sch11}) and (\ref{sch12}) is that according to (\ref{sch12}) the projection of angular momentum onto the direction of radiation is zero due to the definition (\ref{Lds-nn}). In a sense, (\ref{sch12}) does not contain the  spin of the photons. At the same time, the projection of angular momentum onto the direction of radiation, calculated by use of  (\ref{sch11}), generally speaking, is not equal to zero. There is no contradiction in this, because the spatial distribution of the angular momentum density is not uniquely determined. 
A consequence of the Noether's theorem is the conservation of  the integral angular momentum over the entire space. Integrals over $\d\Omega$ in equations (\ref{sch11}) and (\ref{sch12}) give the same result.
A remarkable consequence of both (\ref{sch11}) and (\ref{sch12}) is that radiation at a frequency  $n\omega$ carries an angular momentum equal to $n\hbar$ per photon on average.

Next we show that in the ultrarelativistic limit the asymptotic expressions for  (\ref{sch11}) and (\ref{sch12}) gives the same result.
For $\gamma\gg 1$ in  (\ref{sch11}) and (\ref{sch12}), the maximum in the sums occurs at very large numbers $n$. In this case, the Bessel functions $J_n$ can be expressed through modified Bessel functions of the third kind $K_\nu$
\begin{align}\label{JK}
J_n(nx)\approx &\frac{1}{\pi\sqrt{3}}(1-x^2)^{1/2}K_{1/3}\left[\frac n3(1-x^2)^{3/2}\right],\\
J'_n(nx)\approx &\frac{1}{\pi\sqrt{3}}(1-x^2)K_{2/3}\left[\frac n3(1-x^2)^{3/2}\right].
\end{align}
Summation over $n$ can be replaced by integration over a continuously changing ``harmonic number'' $\nu=\tilde{\omega}/\omega$, where $\tilde\omega$ is the radiation frequency. The procedure for expansion in small quantity $\gamma^{- 1}$, in principle, coincides with the asymptotic expansion of the Schott formula in the theory of synchrotron radiation. Therefore we will present only the final result, referring the reader for details, for example, to the books \cite{SokolovTernov, Hofmann_2004, Bordovitsyn-SR}. 

As a result of this expansion, both  (\ref{sch11}) and (\ref{sch12}) take the form
\be\label{dldt}
\frac{\d L_z}{\d\Omega\d \tilde\omega\d t}=\frac{{2}e^2\tilde\omega^2(1+\psi^2)}{3\pi^2 c\omega^2\gamma^4}\left[\psi^2K_{1/3}^2(\eta)+(1+\psi^2)K_{2/3}^2(\eta)\right].
\ee
The reduced angle $\psi_z=\gamma(\pi/2-\vartheta)$ is measured from the orbital plane. The argument of the Bessel functions is equal to
\[
\eta=\frac{\tilde\omega}{3\omega\gamma^3}(1+\psi_z^2).
\]
The maximum angular momentum flux (\ref{dldt}) lies in the region $\eta\sim 1$, i.e. at frequencies of the order
\[
\tilde\omega\sim \frac{3\omega\gamma^3}{1+\psi_z^2}.
\]

Integrating  (\ref{dldt}) over the solid angle gives the spectral distribution of the total angular momentum flux
\be
\label{dldnu}
\frac{\d L_{z}}{\d \tilde\omega\d t}=\frac{\sqrt{3}\,e^2\gamma}{2\pi c} y\int_y^\infty K_{5/3}(x)\d x,
\ee
where
\[
y=\frac{2\tilde\omega}{3\omega\gamma^3}.
\]

The angular distribution of the angular momentum flux  is obtained by integration of  (\ref{dldt}) over frequency
\be\label{sch4}
\frac{\d L_{z}}{\d\Omega\d t}=\frac{e^2\omega\gamma^{5}}{32\pi c}\frac{7+12\psi_z^2}{(1+\psi_z^2)^{7/2}}.
\ee
The same expression can be obtained if the total angular momentum (\ref{Lsfer3d}) received by the observer during one flash is divided by the period of rotation around the circle $2\pi/\omega$.
The difference in signs is due to the fact that in this section the $z$ axis has the opposite direction.

The flux of angular momentum integrated over angles and frequencies is equal to
\be\label{sch5}
\frac{\d L_{z}}{\d t}=\frac{2e^2\omega\gamma^{4}}{3 c},
\ee
which is consistent with the angular distribution of angular momentum $L'_z$ in  (\ref{i21}), if it is integrated over the angle $\varphi$ and divided by the period of revolution of the particle in a circle.
\section{Conclusion}
The angular momentum of the electromagnetic field  depends on the  coordinate system. We presented the angular momentum of  radiation of an arbitrarily moving charge as the sum of ``intrinsic'' angular momentum $\bm L_0$ calculated relative to the point where the charge is located, and the ``orbital'' angular momentum $\bm L'$ relative to an arbitrary stationary center. In particular, the instantaneous center of curvature of the trajectory is considered as such a center.  The non-zero angular momentum of the radiation relative to the radiation source is due to the fact that the Poynting vector of the radiation field does not coincide with the direction of  radiation. 

The flux of orbital angular momentum  relative to an arbitrary point lying on the principal normal to the trajectory at a distance of $r'$ from the charge,  is $\gamma^2  r'/\rho$ times greater than the flux of intrinsic angular momentum, where $\rho$ is the local radius of curvature of the trajectory at the point of radiation.
Explicit expressions for the angular distribution of angular momentum fluxes $\bm L_0$ and $\bm L'$ for ultrarelativistic charge  are obtained. 

According to the definition, the projection of angular momentum onto the radius vector is zero. This means that radiation absorbed by an infinitesimal object cannot transmit to it a torque  about the direction of  radiation. However, if an object has finite dimensions, it can receive a non-zero angular momentum relative to the radius vector of the center of the object. Since the radiation from an ultrarelativistic charge is concentrated in a narrow cone, a relatively small object of finite size can absorb a significant portion of this radiation. We have shown that the projection of the angular momentum onto the axis of the radiation cone is different from zero, and has opposite signs above and below the trajectory plane.
 We have also shown that in the ultrarelativistic approximation, the canonical angular momentum of the radiation coincides with the angular momentum resulting from the symmetrized energy-momentum tensor of the electromagnetic field.

\section*{Acknowledgements}
This research was supported by  RFBR project No. 19-42-700011.

%

\end{document}